\documentclass[conference]{IEEEtran}

\ifCLASSINFOpdf

\else

\fi

\usepackage{algpseudocode}
\usepackage{algorithm}
\usepackage[dvips]{graphicx}
\usepackage{latexsym}
\usepackage{amssymb}
\usepackage{amsmath}
\usepackage{multirow}
\usepackage{xcolor}
\usepackage{lipsum}
\usepackage{multicol}
\usepackage{enumerate}
\usepackage{algorithm}
\usepackage{algorithmicx}
\usepackage{algpseudocode}
\usepackage{amsmath}
\usepackage{graphicx}
\usepackage{subfig}
\usepackage[noadjust]{cite}

\begin{document}

\title{Hybrid Analog-Digital Precoding for Interference Exploitation}

\author{\IEEEauthorblockN{Ang Li$^1$, Christos Masouros$^1$ and Fan Liu$^{1,2}$}
\IEEEauthorblockA{Dept. of Electronic and Electrical Eng., University College London, London, U.K.$^1$\\
School of Information and Electronics, Beijing Institute of Technology, Beijing 100081, China$^2$ \\
Email: \{ang.li.14, c.masouros\}@ucl.ac.uk$^1$, liufan92@bit.edu.cn$^2$}
}

\IEEEspecialpapernotice{(Invited Paper)}

\maketitle

\begin{abstract}
We study the multi-user massive multiple-input-single-output (MISO) and focus on the downlink systems where the base station (BS) employs hybrid analog-digital precoding with low-cost 1-bit digital-to-analog converters (DACs). In this paper, we propose a hybrid downlink transmission scheme where the analog precoder is formed based on the SVD decomposition. In the digital domain, instead of designing a linear transmit precoding matrix, we directly design the transmit signals by exploiting the concept of constructive interference. The optimization problem is then formulated based on the geometry of the modulation constellations and is shown to be non-convex. We relax the above optimization and show that the relaxed optimization can be transformed into a linear programming that can be efficiently solved. Numerical results validate the superiority of the proposed scheme for the hybrid massive MIMO downlink systems.
\end{abstract}


\IEEEpeerreviewmaketitle

\section{Introduction}
Towards the fifth generation (5G) and future wireless communication standards, the concept of massive multiple-input multiple-output (MIMO) has been introduced in \cite{r11}. With the knowledge of the channel state information (CSI) at the BS, massive MIMO systems can greatly improve the spectral efficiency of the wireless networks. This can be achieved by multi-user transmit precoding, among which it has been shown that linear precoding schemes such as ZF and regularized zero-forcing (RZF) are near-optimal in a fully-digital massive MIMO system \cite{r3}.

Nevertheless, the practical implementation of a fully-digital massive MIMO system may be problematic, due to the significantly increased hardware complexity and resulting power consumption. For practical consideration, one potential technique that can reduce both the hardware complexity and power consumption is to reduce the number of radio frequency (RF) chains by employing the hybrid analog-digital structures \cite{r12}\nocite{r13}\nocite{r14}\nocite{r27}-\cite{r28}, where the precoding is divided into the analog domain and digital domain. In addition to the hybrid precoding, another potential technique is to reduce the power consumption per RF chain by employing very low-resolution digital-to-analog converters (DACs), especially for the 1-bit case. Since the power consumption for DACs grows exponentially with the increasing quantization precision \cite{r15}, the application of 1-bit DACs will significantly reduce the power consumption per RF chain and the resulting total power consumption at the BS. 

Due to the above benefits, transmit beamforming schemes with 1-bit DACs have drawn increasing research attention recently \cite{r16}\nocite{r17}\nocite{r18}\nocite{r19}-\cite{r20}. The performance of 1-bit quantized ZF precoding is analysed in \cite{r16}, while a 1-bit precoding based on the minimum-mean squared error (MMSE) criterion is considered in \cite{r17}. In \cite{r18}-\cite{r20}, non-linear schemes that directly map the data symbols to the transmit signals are proposed, where the precoding method in \cite{r18} is based on the gradient descend method (GDM), an iterative approach based on biconvex relaxation is proposed in \cite{r19}, and several complicated precoding methods based on semidefinite relaxation (SDR) and ${l_\infty }$-norm relaxation are proposed in \cite{r20}. Nevertheless, the above works may not be optimal as it ignores that interference can be exploited on an instantaneous basis \cite{r6}\nocite{r7}\nocite{r8}-\cite{r9}. 

Therefore in this paper, we consider the hybrid transmission scheme for multi-user 1-bit massive MIMO downlink by exploiting the constructive interference (CI). The analog precoder is designed solely based on the CSI to alleviate the complexity of the joint design. In the digital domain, we directly design the quantized signal vector, where the optimization problem is formulated based on the geometry of the modulation constellations. Due to the constraint on the output signals of 1-bit DACs and the power normalization factor, the resulting optimization problem is not convex. We propose a two-step approach, where in the first step we apply a relaxation on the mathematical constraint resulting from the use of 1-bit DACs, such that the resulting optimization is transformed into a linear programming (LP), which can be efficiently solved in polynomial time. We then apply an element-wise normalization on the obtained signal vector to meet the 1-bit DAC transmission. The numerical results show that the proposed scheme can better approach the performance of ideal hybrid systems with infinite-precision DACs. 

$Notations$: $a$, $\bf a$, and $\bf A$ denote scalar, vector and matrix, respectively. ${( \cdot )^T}$, ${( \cdot )^H}$, and $tr\left\{  \cdot  \right\}$ denote transposition, conjugate transposition, and trace of a matrix respectively. $\left|  \cdot  \right|$ denotes the modulus of a complex, and ${{\cal C}^{n \times n}}$ represents an $n \times n$ matrix in the complex set. $\Re ( \cdot )$ and $\Im ( \cdot )$ denote the real and imaginary part of a complex number, respectively.

\section{System Model}
We consider a multi-user 1-bit massive MISO system with the hybrid structure in the downlink, as depicted in Fig.~1. The BS with $N_t$ antennas and $N_{RF}$ RF chains simultaneously serves $K$ single-antenna users on the same time-frequency resource, and ${N_t} \gg N_{RF} \ge K$. Our focus in this paper is the precoding design, and infinite-precision ADCs are assumed for each user. Perfect CSI is also assumed throughout the paper, while the channel estimation for hybrid structures is discussed in \cite{r25} and the references therein. 

\begin{figure}[h]
\centering
\includegraphics[scale=0.25]{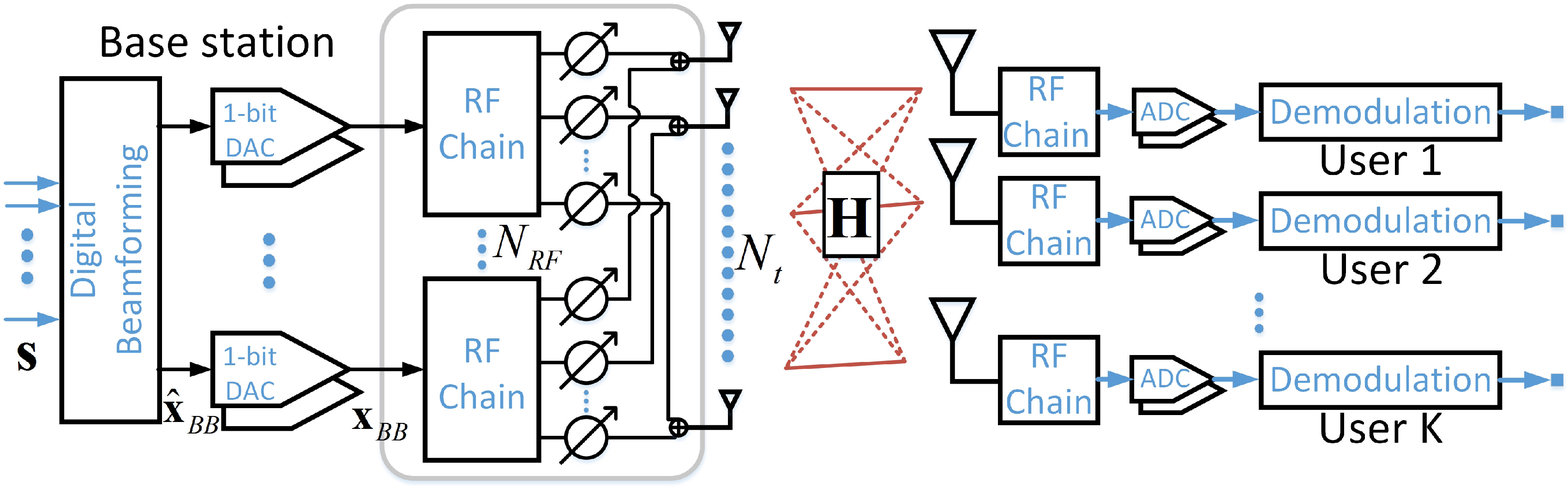}
\caption{Hybrid analog-digital system with 1-bit DACs}
\end{figure}

Following the closely related literature \cite{r16}-\cite{r20}, the symbol vector is assumed to be from a normalized PSK constellation, denoted as ${\bf{s}} \in {{\cal C}^{K \times 1}}$. We denote the unquantized signal vector in the digital domain as ${\bf{\hat x}}_{BB} \in {{\cal C}^{{N_{RF}} \times 1}}$ that is formed based on $\bf s$, expressed as
\begin{equation}
{\bf \hat x}_{BB}={\cal B} \left( {{\bf s}} \right),
\end{equation}
where $\cal B$ denotes a generic linear precoding matrix or a non-linear mapping scheme dependent on $\bf s$. Then, the output signal vector ${\bf x}_{BB}$ of the 1-bit DACs is obtained as
\begin{equation}
{\bf x}_{BB}={\cal Q} \left( {{\bf \hat x}_{BB}} \right),
\end{equation}
where ${\cal Q}$ denotes the element-wise 1-bit quantization in the real and imaginary part of ${\bf \hat x}_{BB}$, and therefore each entry in the resulting ${\bf x}_{BB}$ belongs to the set $\left\{ { \pm \frac{1}{{\sqrt 2 }} \pm \frac{1}{{\sqrt 2 }}j} \right\}$. In the analog domain, we denote ${{\bf{F}}_{RF}} \in {{\cal C}^{{N_t} \times {N_{RF}}}}$ as the analog precoder implemented with analog phase shifters, and therefore each entry of ${\bf F}_{RF}$ is of constant modulus. When a fully-connected RF structure is considered \cite{r12}, as assumed in this paper and shown in Fig.~1, ${\bf F}_{RF}$ can be expressed as
\begin{equation}
{{\bf{F}}_{RF}} = \left[ {{{\bf{f}}_1},{{\bf{f}}_2},...,{{\bf{f}}_{{N_{RF}}}}} \right],
\end{equation}
where each ${{\bf{f}}_k} \in {{\cal C}^{{N_t} \times 1}}$. In this paper, we normalize each entry of ${\bf F}_{RF}$ to satisfy
\begin{equation}
\left| {{{\bf{f}}_k}\left( m \right)} \right| = {1 \mathord{\left/ {\vphantom {1 {{N_t}}}} \right. \kern-\nulldelimiterspace} {{N_t}}}, {\kern 3pt} \forall m \in \left\{ {1,2,...,N_t} \right\}.
\end{equation}
Accordingly, we can express the received signal vector as
\begin{equation}
{\bf{y}} = \sqrt {P}  \cdot {\bf{H}}{{\bf{x}}_T} + {\bf{n}} = \sqrt {P}  \cdot \frac{1}{f} \cdot {\bf{H}}{{\bf{F}}_{RF}}{{\bf{x}}_{BB}} + {\bf{n}},
\end{equation}
where ${\bf{H}} \in {{\cal C}^{K \times {N_t}}}$ denotes the flat-fading Rayleigh channel, and each entry in $\bf H$ follows the standard complex Gaussian distribution. ${\bf n} \in {\cal C}^{K \times 1}$ is the circular symmetric Gaussian distributed additive noise vector with zero mean and covariance $\sigma^2 \cdot {\bf I}$. $P$ denotes the total available transmit power per antenna, and for simplicity we assume a uniform power distribution. ${{\bf{x}}_T} =\frac{1}{f} \cdot {{\bf{F}}_{RF}}{{\bf{x}}_{BB}}$ is the transmit signal vector, and $f$ is the power normalization factor to constrain the signal power after precoding, given by
\begin{equation}
f = \sqrt {tr\left\{ {{{\bf{F}}_{RF}}{{\bf{x}}_{BB}}{\bf{x}}_{BB}^H{\bf{F}}_{RF}^H} \right\}}.
\end{equation}

\section{Hybrid Transmission Scheme based on CI}
For practical consideration, the analog precoder is designed solely dependent on the CSI, which means that the phase shifters only need to change their phases when the channel changes. As for the digital domain, a symbol-level design is required since the output signals are dependent on the data symbols \cite{r19}\cite{r20}. It is this aspect of the transmission that allows us to observe interference from an instantaneous point of view, and exploit it constructively.

\subsection{Analog Precoding Design based on SVD}
We firstly introduce the analog design based on SVD, where we express the SVD of the channel matrix $\bf H$ as
\begin{equation}
{\bf{H}} = {\bf{U}} {\bf \Lambda} {{\bf{V}}^H}.
\end{equation}
In (7), $\bf U$ and ${\bf{V}} = \left[ {{{\bf{v}}_1},{{\bf{v}}_2},...,{{\bf{v}}_{{N_t}}}} \right]$ are unitary matrices that contain the left- and right-singular vectors, and ${\bf{V}} \in {{\cal C}^{{N_t} \times {N_t}}}$. Then, the phases of each phase shifter in the analog precoder ${\bf F}_{RF}$ are selected as the phases in the first $N_{RF}$ columns of $\bf V$, expressed as
\begin{equation}
{{\bf{f}}_k}\left( m \right) = \frac{1}{{{N_t}}}{e^{j \cdot {\phi _m}}}, {\kern 3pt} \forall m \in \left\{ {1,2,...,{N_t}} \right\},
\end{equation}
where $\phi_m$ is the phase of the $m$-th entry in ${\bf v}_k$. 

$Remark$: In addition to the analog precoding scheme based on SVD, other analog schemes such as discrete Fourier transform (DFT) codebooks or matched filtering (MF) \cite{r24} can also be applied upon the digital precoding scheme introduced in the following. Compared to the MF scheme, one of the advantages for our adopted scheme is its applicability in the case of $N_{RF} > K$. 

\subsection{CI-based Digital Precoding}
CI is defined as the interference that pushes the received symbols away from the detection thresholds of the modulation constellation \cite{r6}-\cite{r9}. The exploitation of the CI is firstly introduced in \cite{r6}, while the constructive region has been further introduced in \cite{r9}, where it is shown that, as long as the resulting interfered signals are located in the constructive region, the distance to the decision thresholds is increased. While we focus on the PSK modulation in this paper, the extension to QAM modulations are applicable, and we refer the readers to \cite{r9}\cite{r22} for a detailed description.

Before introducing the precoding design in the digital domain, we first express the equivalent user-to-RF channel as
\begin{equation}
{{\bf{H}}_e} = {\bf{H}}{{\bf{F}}_{RF}},
\end{equation}
where ${{\bf{H}}_e} \in {{\cal C}^{K \times {N_{RF}}}}$, based on which we design the digital scheme. For the digital precoding design, we decompose the equivalent channel ${\bf H}_e$ into
\begin{equation}
{{\bf{H}}_e} = {\left[ {{\bf{h}}_1^T,{\bf{h}}_2^T,...,{\bf{h}}_K^T} \right]^T},
\end{equation}
where each ${\bf h}_k \in {\cal C}^{1 \times N_{RF}}$ represents the equivalent channel of user $k$, and the received signal for user $k$ is then expressed as
\begin{equation}
{y_k} = \sqrt {P}  \cdot \frac{1}{f} \cdot {{\bf{h}}_k}{{\bf{x}}_{BB}} + {n_k},
\end{equation}
where $n_k$ is the $k$-th entry in $\bf n$. Following the symbol-scaling methods introduced in \cite{r26}, we first decompose each data symbol along the corresponding detection thresholds, expressed as
\begin{equation}
{s_k} = s_k^{\cal A}  + s_k^{\cal B},
\end{equation}
where $s_k^{\cal A}$ and $s_k^{\cal B}$ are parallel to the two detection thresholds of $s_k$, respectively. Similarly, by introducing
\begin{equation}
\alpha _k^{\cal A} \ge 0, {\kern 3pt} \alpha _k^{\cal B} \ge 0, {\kern 3pt} \forall k \in \left\{ {1,2, \cdots ,K} \right\},
\end{equation}
we decompose the noiseless received signal along the detection thresholds, given by
\begin{equation}
{{\bf{h}}_k}{{\bf{x}}_{BB}} = \alpha _k^{\cal A} s_k^{\cal A}  + \alpha _k^{\cal B} s_k^{\cal B}.
\end{equation}

An illustrative example is shown in Fig. 2, where we focus on one constellation point of a normalized 8PSK constellation. Without loss of generality, we assume $\mathop {OS}\limits^ \to   = {s_k}$ is the data symbol for user $k$ and $\mathop {OB}\limits^ \to   = {{\bf{h}}_k}{{\bf{x}}_{BB}}$ denotes the noiseless received signal for user $k$. We then decompose both $\mathop {OS}\limits^ \to$ and $\mathop {OB}\limits^ \to$ along the two detection thresholds $\mathop {OD}\limits^ \to$ $\left({y_k^{\cal A}}\right)$ and $\mathop {OE}\limits^ \to$ $\left( {y_k^{\cal B}} \right)$ of the data symbol $s_k$. It is then obvious that the performance is dependent on the values of each $\alpha_k^{\cal A}$ and $\alpha _k^{\cal B}$, and a larger value of $\alpha_k^{\cal A}$ and $\alpha _k^{\cal B}$ represents a larger distance to the detection thresholds.

Accordingly, we propose to maximize the minimum value of $\alpha_k^{\cal U}$, ${\cal U} \in \left\{ {{\cal A}, {\cal B}} \right\}$ such that the received signals are pushed as far as possible away from the detection thresholds. For a 1-bit downlink transmission, this leads to the following optimization problem
\begin{equation}
\begin{aligned}
&\mathcal{P}_1: {\kern 3pt} \mathop {\max }\limits_{{{\bf{x}}_{BB}}} \mathop {\min }\limits_{k, {\kern 1pt} {\cal U}} {\kern 3pt} \frac{{\alpha _k^{\cal U}}}{f} \\
&{\kern 0pt} s. t. {\kern 10pt} {{\bf{h}}_k}{{\bf{x}}_{BB}} = \alpha _k^{\cal A} s_k^{\cal A}  + \alpha _k^{\cal B} s_k^{\cal B}, {\kern 3pt} \forall k \in {\cal K}, {\kern 2pt} {\cal U} \in \left\{ {{\cal A}, {\cal B}}\right\}\\
&{\kern 24pt} {x_n} \in \left\{ { \pm \frac{1}{{\sqrt 2 }} \pm \frac{1}{{\sqrt 2 }}j} \right\}, {\kern 3pt} \forall n \in {\cal N}\\
&{\kern 24pt} f = \sqrt {tr\left\{ {{{\bf{F}}_{RF}}{{\bf{x}}_{BB}}{\bf{x}}_{BB}^H{\bf{F}}_{RF}^H} \right\}}
\end{aligned}
\end{equation}
where $x_n$ is the $n$-th entry in ${\bf x}_{BB}$, ${\cal K} = \left\{ {1,2,...,K} \right\}$ and ${\cal N} = \left\{ {1,2,...,{N_{RF}}} \right\}$. The above optimization problem ${\cal P}_1$ is non-convex and difficult to solve because of the following reasons. Firstly, each $x_n$ is constrained to specific values due to the deployment of the 1-bit DACs. Moreover, the optimization variable ${\bf x}_{BB}$ is included in the expression of the scaling factor. To further simplify the above non-convex problem, we employ the following relaxation.

\begin{figure}[!t]
\centering
\includegraphics[scale=0.3]{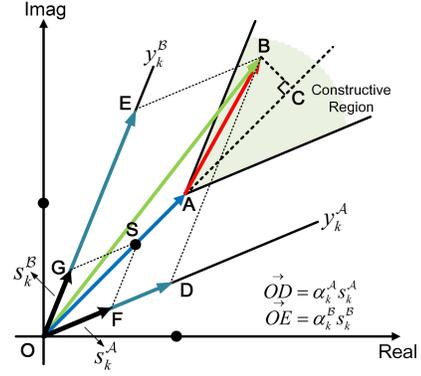} 
\caption{Signal decomposition based on CI for QPSK}
\end{figure}  

\subsubsection{Relaxation}
We first simplify the objective function of ${\cal P}_1$ to get rid of the effect of the scaling factor $f$, which leads to
\begin{equation}
{\cal F}\left( {{{\cal P}_1}} \right) = \mathop {\max }\limits_{{{\bf{x}}_{BB}}} \mathop {\min }\limits_{k, {\kern 1pt} {\cal U}} {\kern 3pt} {{\alpha _k^{\cal U}}}.
\end{equation}
While the above transformation may result in sub-optimal results, it greatly simplifies the formulation and further enables the efficient solutions. Moreover, it will be shown in the results that the obtained solutions achieve a near-optimal performance. After this transformation, the resulting optimization is still non-convex due to the constraint on the output signals of 1-bit DACs. Then, we further relax the strict modulus constraint for each $x_n$ on its real and imaginary part respectively, expressed as
\begin{equation}
\left| {\Re \left( {{\hat x_n}} \right)} \right| \le \frac{1}{{\sqrt 2 }}, {\kern 3pt} \left| {\Im \left( {{\hat x_n}} \right)} \right| \le \frac{1}{{\sqrt 2 }}, {\kern 3pt} \forall n \in {\cal N}.
\end{equation}
Accordingly, we can reformulate the optimization problem ${\cal P}_1$ into its relaxation form ${\cal P}_2$ as
\begin{equation}
\begin{aligned}
&\mathcal{P}_2: {\kern 3pt} \mathop {\min }\limits_{{{\bf \hat {x}}_{BB}}} {\kern 3pt} -t \\
&{\kern 0pt} s. t. {\kern 10pt} {{\bf{h}}_k}{{\bf \hat {x}}_{BB}} = \alpha _k^{\cal A} s_k^{\cal A}  + \alpha _k^{\cal B} s_k^{\cal B}, {\kern 3pt} \forall k \in {\cal K}\\
&{\kern 24pt} t - {\alpha _k}^{\cal U} \le 0, {\kern 3pt} {\cal U} \in \left\{ {{\cal A}, {\cal B}}\right\}, {\kern 3pt} \forall k \in {\cal K}\\
&{\kern 22pt} \left| {\Re \left( {\hat x}_n \right)} \right| \le \frac{1}{{\sqrt 2 }}, {\kern 3pt} \left| {\Im \left( {\hat x}_n \right)} \right| \le \frac{1}{{\sqrt 2 }}, {\kern 3pt} \forall n \in {\cal N}
\end{aligned}
\end{equation}
where we denote the relaxed signal vector and its entry obtained by ${\cal P}_2$ as ${\bf \hat x}_{BB}$ and ${\hat x}_n$, respectively. In the following we further show that ${\cal P}_2$ can be transformed as a LP optimization, which can be efficiently solved in polynomial time.

\subsubsection{LP Transformation}
To obtain the LP formulation, we firstly expand (14) as
\begin{equation}
\left[ {\begin{array}{*{20}{c}}
{\Re \left( {{{\bf{h}}_k}} \right)}&{ - \Im \left( {{{\bf{h}}_k}} \right)}\\
{\Im \left( {{{\bf{h}}_k}} \right)}&{\Re \left( {{{\bf{h}}_k}} \right)}
\end{array}} \right]\left[ {\begin{array}{*{20}{c}}
{{\bf \hat {x}}_{BB}^\Re }\\ {{\bf \hat {x}}_{BB}^\Im } \end{array}} \right] = \alpha _k^\Re s_k^\Re  + \alpha _k^\Im s_k^\Im,
\end{equation}
where ${\bf \hat x}_{BB}^{\Re}$ and ${\bf \hat x}_{BB}^{\Im}$ denote the real and imaginary part of ${\bf \hat x}_{BB}$. Based on the signal decomposition in \cite{r26}, we obtain the expression of $\alpha _k^\Re$ and $\alpha _k^\Im$, given by
\begin{equation}
\begin{aligned}
\alpha _k^{\cal A} &= \frac{{B_k^\Im {\bf{h}}_k^\Re  - B_k^\Re {\bf{h}}_k^\Im }}{A_k^\Re B_k^\Im  - A_k^\Im B_k^\Re}{\bf{\hat x}}_{BB}^\Re  - \frac{{B_k^\Im {\bf{h}}_k^\Im  + B_k^\Re {\bf{h}}_k^\Re }}{A_k^\Re B_k^\Im  - A_k^\Im B_k^\Re}{\bf{\hat x}}_{BB}^\Im, \\
\alpha _k^{\cal B}& = \frac{{A_k^\Re {\bf{h}}_k^\Im  - A_k^\Im {\bf{h}}_k^\Re }}{A_k^\Re B_k^\Im  - A_k^\Im B_k^\Re}{\bf{\hat x}}_{BB}^\Re  + \frac{{A_k^\Re {\bf{h}}_k^\Re  + A_k^\Im {\bf{h}}_k^\Im }}{A_k^\Re B_k^\Im  - A_k^\Im B_k^\Re}{\bf{\hat x}}_{BB}^\Im,
\end{aligned}
\end{equation}
where we express
\begin{equation}
s_k^{\cal A} = A_k^\Re  + j \cdot A_k^\Im, {\kern 5pt} s_k^{\cal B} = B_k^\Re  + j \cdot B_k^\Im, {\kern 3pt} \forall k \in {\cal K}.
\end{equation}
In (20), for simplicity of notations we have introduced ${\bf{h}}_k^\Re  = \Re \left( {{{\bf{h}}_k}} \right)$ and ${\bf{h}}_k^\Im  = \Im \left( {{{\bf{h}}_k}} \right)$. By further defining
\begin{equation}
\begin{aligned}
&{{\bf{A}}_k} = \frac{{B_k^\Im {\bf{h}}_k^\Re  - B_k^\Re {\bf{h}}_k^\Im }}{{A_k^\Re B_k^\Im  - A_k^\Im B_k^\Re }}, {\kern 5pt} {{\bf{B}}_k} =  - \frac{{B_k^\Im {\bf{h}}_k^\Im  + B_k^\Re {\bf{h}}_k^\Re }}{{A_k^\Re B_k^\Im  - A_k^\Im B_k^\Re }},\\
&{{\bf{C}}_k} = \frac{{A_k^\Re {\bf{h}}_k^\Im  - A_k^\Im {\bf{h}}_k^\Re }}{{A_k^\Re B_k^\Im  - A_k^\Im B_k^\Re }}, {\kern 6pt} {{\bf{D}}_k} = \frac{{A_k^\Re {\bf{h}}_k^\Re  + A_k^\Im {\bf{h}}_k^\Im }}{{A_k^\Re B_k^\Im  - A_k^\Im B_k^\Re }},
\end{aligned}
\end{equation}
and
\begin{equation}
\begin{aligned}
&{{\bf{R}}_k} = \left[ {\begin{array}{*{20}{c}}
{{{\bf{A}}_k}}&{{{\bf{B}}_k}}
\end{array}} \right], {\kern 3pt} {{\bf{I}}_k} = \left[ {\begin{array}{*{20}{c}}
{{{\bf{C}}_k}}&{{{\bf{D}}_k}}
\end{array}} \right] \\
&{\bf{M}}_0 = {\left[ {{\bf{R}}_1^T,{\bf{R}}_2^T, \cdots, {\bf{R}}_K^T,{\bf{I}}_1^T,{\bf{I}}_2^T, \cdots ,{\bf{I}}_K^T} \right]^T}
\end{aligned}
\end{equation}
(20) can be transformed into a matrix form as
\begin{equation}
{\bf \Lambda}_0  = {\bf{M}}_0{\bf \hat {x}}_0,
\end{equation}
where ${\bf \hat{x}}_0 = {\left[ {{{\left( {{\bf \hat{x}}_{BB}^\Re } \right)}^T},{{\left( {{\bf \hat{x}}_{BB}^\Im } \right)}^T}} \right]^T}$ and
\begin{equation}
{\bf \Lambda}_0  = {\left[ {\alpha _1^{\cal A} , \alpha _2^{\cal A} , \cdots ,\alpha _K^{\cal A} ,\alpha _1^{\cal B} , \alpha _2^{\cal B} ,\cdots ,\alpha _K^{\cal B} } \right]^T}.
\end{equation}
By stacking $t$ into the variable vector and defining
\begin{equation}
{\bf \hat{x}} = {\left[ {\begin{array}{*{20}{c}}
t&{{\bf \hat{x}}_0^T}
\end{array}} \right]^T}, {\kern 2pt} {\bf \Lambda}  = {\left[ {\begin{array}{*{20}{c}}
t&{{\bf \Lambda} _0^T}
\end{array}} \right]^T}, {\kern 2pt} {\bf{M}} = \left[ {\begin{array}{*{20}{c}}
1&{\bf{0}}\\
{\bf{0}}&{{{\bf{M}}_0}}
\end{array}} \right],
\end{equation}
(24) can be further transformed into ${\bf \Lambda}  = {\bf{M}}{\bf \hat {x}}$ where ${\bf M} \in {\cal C}^{(2K+1)\times(2N_{RF}+1)}$. We further introduce a matrix ${\bf T} \in {\cal C}^{2K \times (2K+1)}$
\begin{equation}
{\bf{T}} = \left[ {\begin{array}{*{20}{c}}
{\bf{1}}&{ - {\bf{I}}}
\end{array}} \right],
\end{equation}
where ${\bf{1}} = {\left[ {1,1, \cdots ,1} \right]^T}$. Accordingly, the constraint $t - {\alpha _k}^{\cal U} \le 0$, $\forall k \in {\cal K}$, ${\cal U}=\left\{ {{\cal A}, {\cal B}}\right\}$ can be transformed into a matrix form as
\begin{equation}
{\bf{TM}}{\bf \hat x} \le {\bf{0}}.
\end{equation}
Finally, ${\cal P}_2$ can be transformed into a LP, given by
\begin{equation}
\begin{aligned}
&\mathcal{P}_3: {\kern 3pt} \mathop {\min }\limits_{{{\bf \hat {x}}}} {\kern 3pt} -t \\
&{\kern 0pt} s. t. {\kern 10pt} {\bf{TM}}{\bf \hat x} \le {\bf{0}} \\
&{\kern 21pt}  - \frac{1}{{\sqrt 2 }} \le {\hat x_E^n} \le \frac{1}{{\sqrt 2 }}, \forall n \in \left\{ {2,3, \cdots ,2{N_{RF}}+1} \right\}
\end{aligned}
\end{equation}
where we note that the constraint in (14) has already been included implicitly with (19)-(28). In ${\cal P}_3$, $\hat x_E^n$ denotes the $n$-th entry in $\bf \hat x$ defined in (26), and we note $\hat x_E^1=t$. Thanks to the hybrid structure, the size of ${\bf \hat x}$ is reduced from $(2N_t+1) \times 1$ to $(2N_{RF}+1) \times 1$, and ${\cal P}_3$ can be efficiently solved in polynomial time. Subsequently, ${\bf \hat x}_{BB}$ can be obtained based on ${\bf \hat x}_0$, expressed as
\begin{equation}
{{\bf{\hat x}}_{BB}} = {\bf{U}}{{\bf{\hat x}}_0},
\end{equation}
where the transformation matrix ${\bf{U}} = \left[ {\begin{array}{*{20}{c}}
{\bf{1}}&{j \cdot {\bf{1}}}\end{array}} \right]$.

\subsubsection{1-Bit Normalization}
Since the above solution may not always guarantee the 1-bit transmission, each ${\hat x}_n$ in ${\bf \hat x}_{BB}$ is further normalized according to
\begin{equation}
{x_n} = \frac{{{\mathop{\rm sgn}} \left[ {\Re \left( {{{\hat x}_n}} \right)} \right]}}{{\sqrt 2 }} + j \cdot \frac{{{\mathop{\rm sgn}} \left[ {\Im \left( {{{\hat x}_n}} \right)} \right]}}{{\sqrt 2 }}, {\kern 3pt} n \in {\cal N},
\end{equation}
where ${\mathop{\rm sgn}} \left[  \cdot  \right]$ is the sign function. The above element-wise normalization guarantees that the 1-bit DAC transmission is met.

\section{Numerical Results}
In this section, we present the numerical results of the proposed scheme based on Monte Carlo simulations. The transmit signal-to-noise ratio (SNR) in each plot is defined as $\rho  = {P \mathord{\left/ {\vphantom {1 {{\sigma ^2}}}} \right.\kern-\nulldelimiterspace} {{\sigma ^2}}}$. For fairness of comparison, existing schemes with 1-bit DACs \cite{r16}-\cite{r20} are applied with the hybrid structure. We compare our proposed scheme (`CI Hybrid 1-Bit') with existing linear precoding approaches with 1-bit DACs (`ZF Hybrid 1-Bit' scheme in \cite{r16}, the MMSE-based scheme `WFQ Hybrid 1-Bit' in \cite{r17} and `GP Hybrid 1-Bit' in \cite{r18}), and the non-linear schemes (`Pokemon Hybrid 1-Bit' in \cite{r19} and `SQUID Hybrid 1-Bit' in \cite{r20}). We denote the fully-digital ZF scheme and hybrid ZF scheme with ideal DACs as `ZF FD' and `ZF Hybrid Ideal'. Both QPSK and 8PSK modulations are considered in the simulations.

\begin{figure}[!b]
\centering
\includegraphics[scale=0.4]{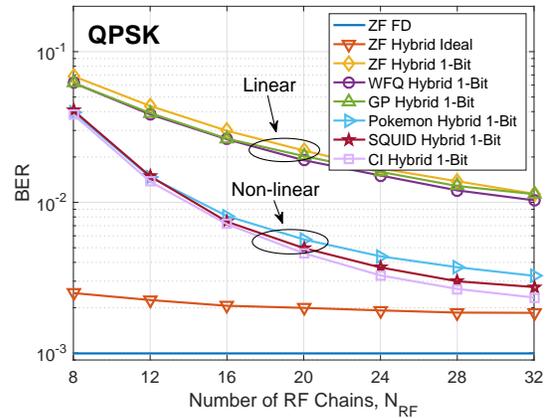}
\caption{BER performance v.s. number of RF chains $N_{RF}$, $N_t$=128, $K$=4, SNR=-5dB, QPSK}
\end{figure}

In Fig.~3, we show the BER performance of the proposed scheme with respect to the increasing number of RF chains. The number of RF chains does not significantly affect the performance of hybrid ZF with ideal DACs, while all hybrid beamforming schemes with 1-bit DACs achieve an improved performance with the increase in the number of RF chains. We have also observed that the proposed scheme `CI Hybrid 1-Bit' greatly outperforms existing quantized linear precoding schemes and is also superior to the non-linear `Pokemon Hybrid 1-Bit' and `SQUID Hybrid 1-Bit'. It's also observed that hybrid schemes with 1-bit DACs require a larger number of RF chains to achieve a close-to-optimal performance.

\begin{figure}[!t]
\centering
\includegraphics[scale=0.4]{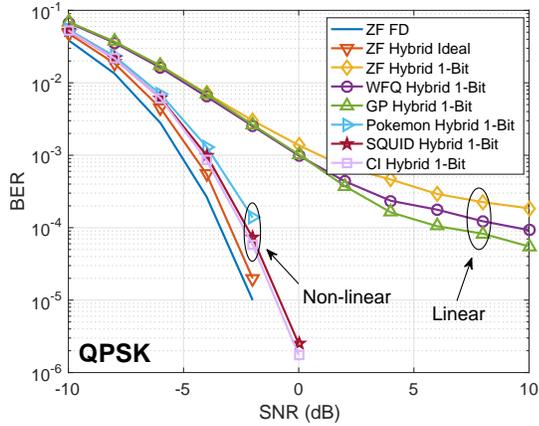}
\caption{BER performance v.s. transmit SNR $\rho$, $N_t$=128, $K$=4, $N_{RF}$=32, QPSK}
\end{figure}

In Fig.~4 and Fig.~5, we show the BER performance with respect to the increasing SNR with $N_{RF}=32$ RF chains for QPSK and 8PSK, respectively. In both Fig.~4 and Fig.~5, it can be observed that the proposed scheme based on constructive interference achieves an improved performance over existing linear and non-linear schemes for both QPSK and 8PSK, while we highlight that the proposed scheme is based on LP optimization that can be efficiently solved in polynomial time, which reveals the superiority of the proposed scheme.

\begin{figure}[!t]
\centering
\includegraphics[scale=0.4]{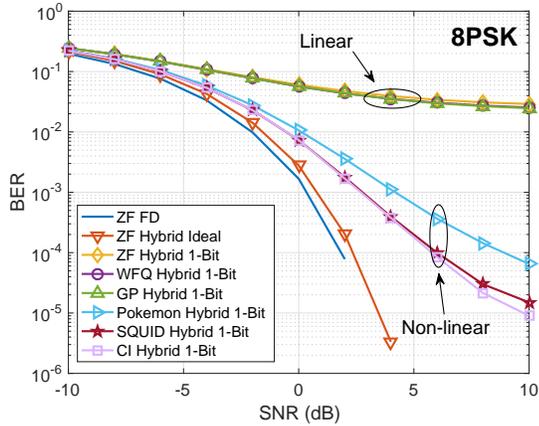}
\caption{BER performance v.s. transmit SNR $\rho$, $N_t$=128, $K$=4, $N_{RF}$=32, 8PSK}
\end{figure}

\section{Conclusion}
In this paper, we propose a CI-based hybrid transmission scheme for multi-user massive MIMO downlink systems with 1-bit DACs at the BS. We employ the right-singular vectors of the channel as the analog precoder, while in the digital domain we directly optimize the output signals of the DACs based on the CI concept. A two-step approach is then proposed to solve the non-convex optimization problem. It is shown in the simulation results that the proposed method outperforms existing precoding approaches designed for downlink massive MIMO systems with 1-bit DACs, and the performance gain is more significant when the ratio of the number of RF chains to users is large.

\section*{Acknowledgment}
This work was supported by the Royal Academy of Engineering, U.K., the Engineering and Physical Sciences Research Council (EPSRC) project EP/M014150/1, and the China Scholarship Council (CSC).

\bibliographystyle{IEEEtran}
\bibliography{refs.bib}

\end{document}